\documentstyle[aps,manuscript,epsf]{revtex} 

\begin{document}
\title{\bf 
                Local Distributions and Rate Fluctuations
                 in a Unified Scaling Law for Earthquakes
}
\author{
\'Alvaro Corral%
\cite{email} 
}
\address{
Departament de F\'\i sica, 
Universitat Aut\`onoma de Barcelona,
Edifici Cc, E-08193 Bellaterra, Barcelona, Spain 
}
%
\date{\today}

\maketitle 
%
%
\begin{abstract}
A recently proposed unified scaling law for interoccurrence times of
earthquakes is analyzed,
both theoretically and with data from Southern California.
We decompose the corresponding probability density
into local-instantaneous distributions,
which scale with the rate of earthquake occurrence.
The fluctuations 
of the rate,
characterizing the non-stationarity of the process,
show a doubly power-law distribution 
and are fundamental to determine the overall behavior,
described by a double power law as well.
\end{abstract}
%

%

\narrowtext
%
%
\newpage   


Earthquakes constitute an extremely complex phenomenon in nature,
with the deformation and sudden rupture of some parts of the Earth crust
driven by
convective motion in the mantle,
and the radiation of energy in the form of seismic waves.
Only a part of this complexity is collected by earthquake catalogs,
where magnitude, epicenter spatial coordinates, and starting time
of events, among other measurements, are recorded.
This information, which converts the phenomenon in a spatio-temporal
point process marked by the magnitude, 
nevertheless 
reveals some important scale-invariant properties.

First, the Gutenberg-Richter law states that the number of earthquakes
in some region with magnitude larger than some threshold value
decreases exponentially with the threshold.
Taking into account that (to a first approximation)
the released energy increases exponentially with the magnitude,
the probability distribution of the released energy turns out to be a power law,
precisely the hallmark of scale-free behavior \cite{Gutenberg,Turcotte,Kagan}.
Second, the introduction of fractal geometry soon led to the recognition
that the spatial distribution of epicenters (or hypocenters)
draws a fractal object over the Earth surface \cite{Turcotte,Kagan}.
And third, the Omori law, proposed more than 100 years ago,
accounts for the number of events (called aftershocks) 
that follow a large shock after some time.
This number is another power law, with exponent close to minus one
\cite{Omori}.


This lack of characteristic scales suggests that the crust 
is in a critical state, 
like the well-known critical points studied in equilibrium systems,
but without external adjustments of control parameters.
Therefore, one may talk about a self-organized critical (SOC) state
for the seismic system \cite{Bak}.
This concept has important implications for the issue of earthquake
prediction \cite{Main};
indeed, a critical crust implies that a fracture process may 
or may not develop to provoke a large earthquake
depending on minor microscopic details
that are intrinsically out of control.

An important quantity characterizing earthquake occurrence
is the time interval between successive earthquakes.
This time (that can be referred to as interoccurrence time,
recurrence time, or waiting time)
although related to the Omori law,
has a distribution that is not clearly known.
In fact, all the possibilities have been proposed, 
from periodic behavior for large earthquakes 
to totally random occurrence.
The most extended view is to consider the existence of two
separated processes, 
one for the main shocks, which should occur randomly following a 
Poisson distribution,
and another process for the aftershocks; 
but this should not hold for large events, for which clustering
has been reported \cite{times}.
In general, the usual studies proceed by fixing 
a limited area of observation where 
aftershocks are skilfully identified and removed from data.
On the opposite side, other works concentrate only 
on series of aftershocks.


Bak, Christensen, {\it et al.} \cite{BCDS} 
have followed an alternative approach,
which is to consider the problem in its complete spatio-temporal
complexity.
They divide the area of South California
into regions of size $L$ degrees in the north-south (meridian) direction
and $L$ degrees as well in the east-west (parallel) direction \cite{L}.
Only earthquakes with magnitude $m$ larger than a threshold value $m_c$
are taken into account
(but no other events are eliminated,
all shocks are equally treated).
For each $L \times L-$region the time interval $\tau$ 
between consecutive earthquakes
is obtained for 
the period from 1984 to 2000
as $\tau_i=t_{i}-t_{i-1}$,
where $t_i$ is the time coordinate of the $i-$th earthquake
within the region with $m > m_c$.
The probability density for this interoccurrence time, 
$D(\tau,m_c,L)$, is computed and the results give 
$
     D(\tau,m_c,L) \propto 1/\tau \ \mbox{ for short times }
$
and a faster decay for long times,
with a dependence also on $L$ and $m_c$.

Remarkably, when a scaling analysis is performed, 
all the distribution functions corresponding to different values
of $L$ and $m_c$ collapse into a single curve if the axis 
are rescaled by $S^b/L^{d_f}$, with $d_f \simeq 1.2$, $b \simeq 1$,
and $S \equiv 10^{m_c}$
(related to the energy roughly as $S \propto E^{2/3}$).
In mathematical words,
%
\begin{equation}
     D(\tau,m_c,L) \simeq
     \frac{L^{d_f}}{S^b} F \left( \frac{L^{d_f}}{S^b}\tau \right)=
     \frac{1}{\tau} G \left( \frac{L^{d_f}}{S^b}\tau \right);
\end{equation}
this scaling law constitutes the Bak-Christensen-Danon-Scanlon
(BCDS) proposal \cite{BCDS}.
For short times,
the function $G$ shows a slow variation not affecting 
the power-law ($1/\tau$) behavior; 
for long times, a fast decay is obtained,
which could be consistent with an exponential distribution 
and therefore with a Poisson process,
according to Ref. \cite{BCDS}.

(From now on, to simplify the notation, 
we will omit the dependence of $D$ on $L$ and $m_c$ 
and just write $D(\tau)$.)

This result is relevant for several reasons,
among them:
it shows scaling in the spatio-temporal occurrence of earthquakes,
a key element to consider earthquakes as a critical phenomenon.
Second, 
it is the first law that relates interoccurrence times,
the Gutenberg-Richter law (factor $1/S^b$),
and the fractal dimension of the spatial distribution of events ($d_f)$,
allowing a unified description.
Third,
the law is valid for all earthquakes, no matter their size or location,
and no matter also if they are considered as aftershocks, foreshocks,
or main shocks.
Fourth, 
the power law tell us that immediately after any earthquake there is a high
probability of having another one,
and this probability decreases in time with no characteristic
scale up to $S^b/L^{d_f}$;
that is, there is a correlation time that depends on the region size
and magnitude under consideration,
and therefore for any event one may find clusters of
aftershocks in all time scales up to an appropriate length scale $L$.

The importance of this law deserves further study.
Here we are interested in a general understanding of the BCDS law
and its origins.
We will analyze the same catalogs as Bak {\it et al.} \cite{catalog}
and will show that the fast decay for long times is not exponential,
but another power law.
$D(\tau)$ is related to 
its local and instant components and to
the rate of earthquake occurrence $r$;
this quantity, which counts the number of events per unit time
in a given region, 
displays large fluctuations across several orders of magnitude,
doubly power-law distributed.
This is in contrast to simple SOC models.

Let us pay more attention to the obtaining
of the distribution $D(\tau)$ by Bak {\it et al.}
As we have mentioned, this distribution accounts for the time difference
between successive earthquakes with magnitude larger than $m_c$
in every $L \times L-$region. 
Times from different regions are counted together in $D(\tau)$.
But the total number of earthquakes differs from region
to region (as it is well-know, due to the fractal spatial distribution), 
with a high variability \cite{variability}.
Therefore, the local distributions $D_{xy}(\tau)$ accounting
for the time difference in a given $L \times L-$region 
(of spatial coordinates $x,y$) are clearly different.
This means that $D(\tau)$ is a mixed distribution constructed from all
the different $D_{xy}(\tau)$.

\begin{figure}
\hspace{15em}
\epsfxsize=3.5truein \hskip 4truein\epsffile{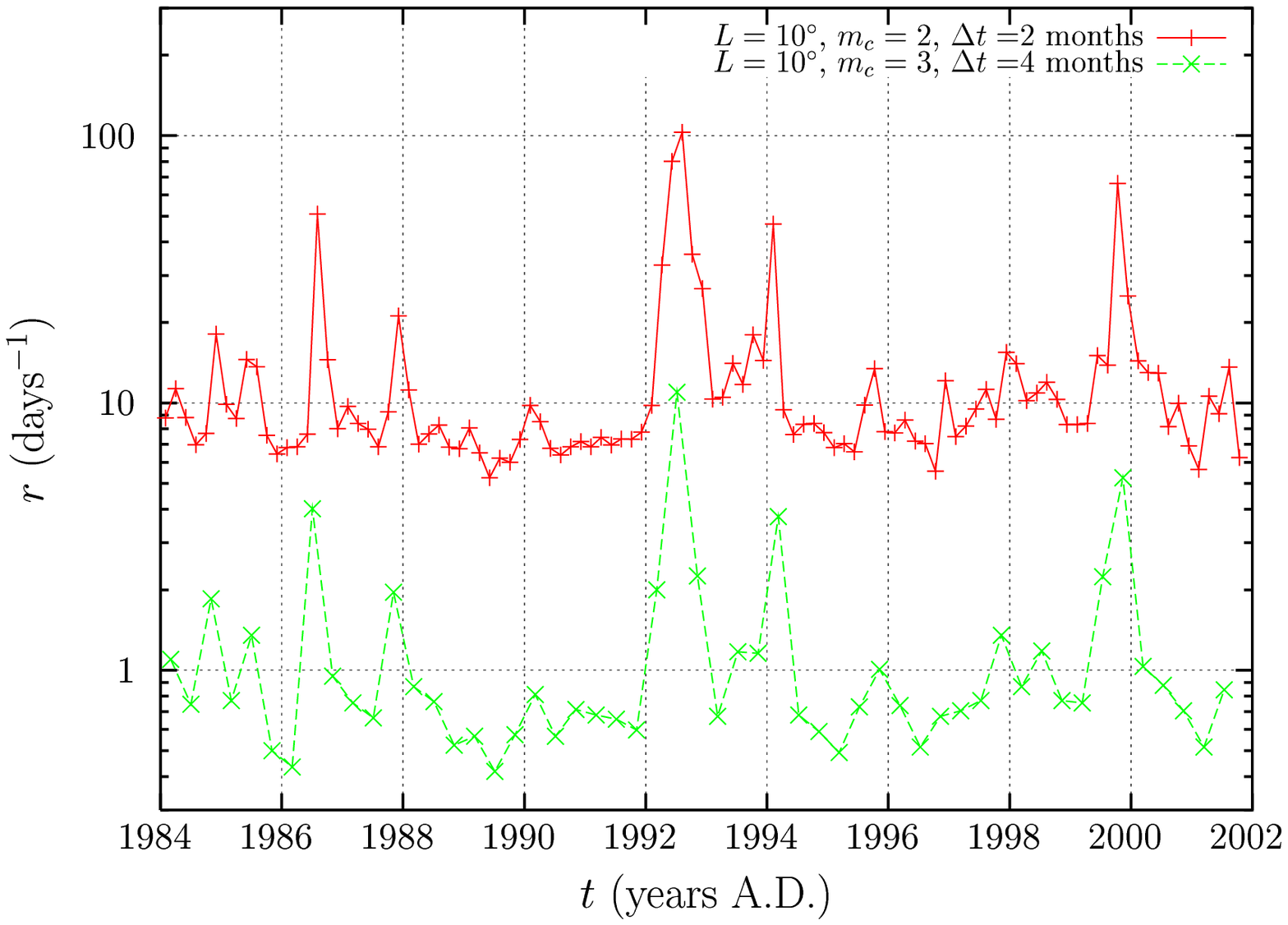} 
\caption{ 
Rate of earthquake occurrence as a function of time
in the $L=10^\circ$ region of South California,
for $m_c=2$ with $\Delta t=2$ months
and for $m_c=3$ with $\Delta t=4$ months.
The vertical log-scale should not make underrate the large
variations in $r$. 
Notice for example the constant rate at 1991 in contrast to 1992.
\label{rate}
}
\end{figure}

But further, looking into a single $L \times L-$region
one can see a high variability in the rate along time $t$
\cite{variability}, see Fig. \ref{rate}.
In fact, the rate typically exhibits a quite stable behavior for some periods
of time, with small fluctuations, but for other periods develops sudden
burst of activity where its value increases sharply and then decreases
to become stationary again, or not.
This intermittency, of course related to the occurrence
of larger earthquakes in the region,
recalls the punctuated-equilibrium behavior of SOC systems \cite{Bak,Gould},
but note however that the variable that displays punctuated equilibrium
is not only the signal $m(t)$, but also the rate.

Again then, the local distribution $D_{xy}(\tau)$ is obtained as a mixture 
of distributions, 
the densities of interoccurrence times in a given region at a certain time $t$,
$D_{xyt}(\tau)$.
From this, we can write
\begin{equation}
     D(\tau) \propto \sum_{\forall x, y} \int D_{xyt}(\tau) r(x,y,t) d t
\end{equation}
where the rate $r(x,y,t)$ is the number of earthquakes per unit time
in the region $(x,y)$ at time $t$.
(All functions here depend as well on $L$ and $m_c$,
though the dependence is not explicitly written.)
The rate $r$ in the integral acts as a weight factor, due to the fact
that the higher the rate in a given region and time, 
the larger the number of earthquakes that are produced and 
contribute to the distribution.

We now make the hypothesis that the dependence on space and time
enters into the distribution $D_{xyt}$ 
only through the rate $r(x,y,t)$. 
That is, we assume that different regions at different times
but with the same rate of occurrence will have the same distribution
of interoccurrence times (if the rate is stationary), i.e.,
\begin{equation}
     D_{xyt}(\tau)=D(\tau|r(x,y,t)),
\end{equation}
which is a conditional density.
Therefore,
\begin{equation}
     D(\tau)=\int_0^{\infty} D(\tau|r) \frac{r \rho(r)}{\mu} dr,
\label{int}
\end{equation}
%
with $\rho(r)$ the probability density of the rate and
$\mu=\langle r \rangle$ just a normalization factor.

\begin{figure}
\epsfxsize=3.5truein \hskip 1truein\epsffile{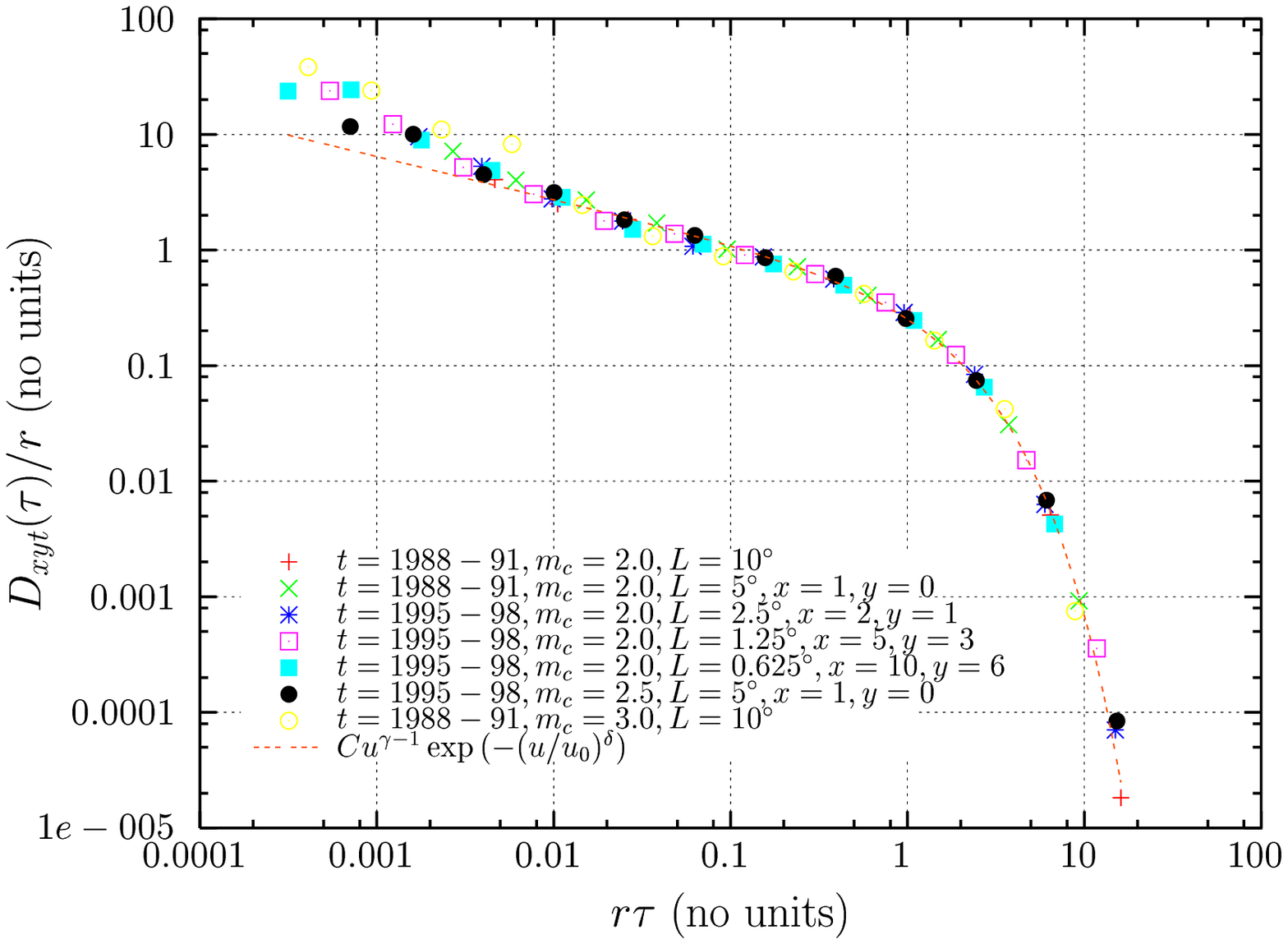} 
\caption{ 
Local distributions of interoccurrence times for several stationary
periods and different regions, after scaling by the rate.
The regions are labeled from $x,y=0$ to $10^\circ/L - 1$
from west to east ($x$) and from south to north ($y$).
The fit is explained in the text;
deviations at small times should be due to short scale
disturbances of the stationarity. 
As in Bak {\it et al.}'s paper, 
times smaller than 38 s are not considered.
\label{Dxy}
}
\end{figure}

Since $D_{xyt}(\tau)$ is an instantaneous quantity (and we have
a single realization of the process), 
it were impossible to measure if we would not have the periods
of stationarity in $r$.
Figure \ref{Dxy} shows these distributions for several periods
of stationarity and several regions of different $L$ and spatial coordinates.
Indeed, the distributions $D_{xyt}$ do not only depend exclusively
on $r$, but they scale with it, i.e.,
\begin{equation}
    D(\tau|r) \simeq r f(r \tau) 
\label{conditional}
\end{equation}
(a behavior that could have been derived by dimensional analysis),
with the scaling function $f$ being a power law for short
times and having a fast decay for large ones.
In fact, the distributions can be fitted by a function of the type
\begin{equation}
    f(u) = 
           C \frac{1}{u^{1-\gamma}}
           e^{-(u/u_0)^\delta},
\label{gammagen}
\end{equation}
with 
$\gamma\simeq 0.63$, $\delta \simeq 0.92$, 
$u_0 \simeq 1.5$, and $C \simeq 0.5$.
We could approximate then $D(\tau|r)$ to a gamma distribution ($\delta=1$),
which ensures that the large-scale cutoff is close to exponential.

\begin{figure}
\hspace{5em}
\epsfxsize=3.5truein \hskip 0.15truein\epsffile{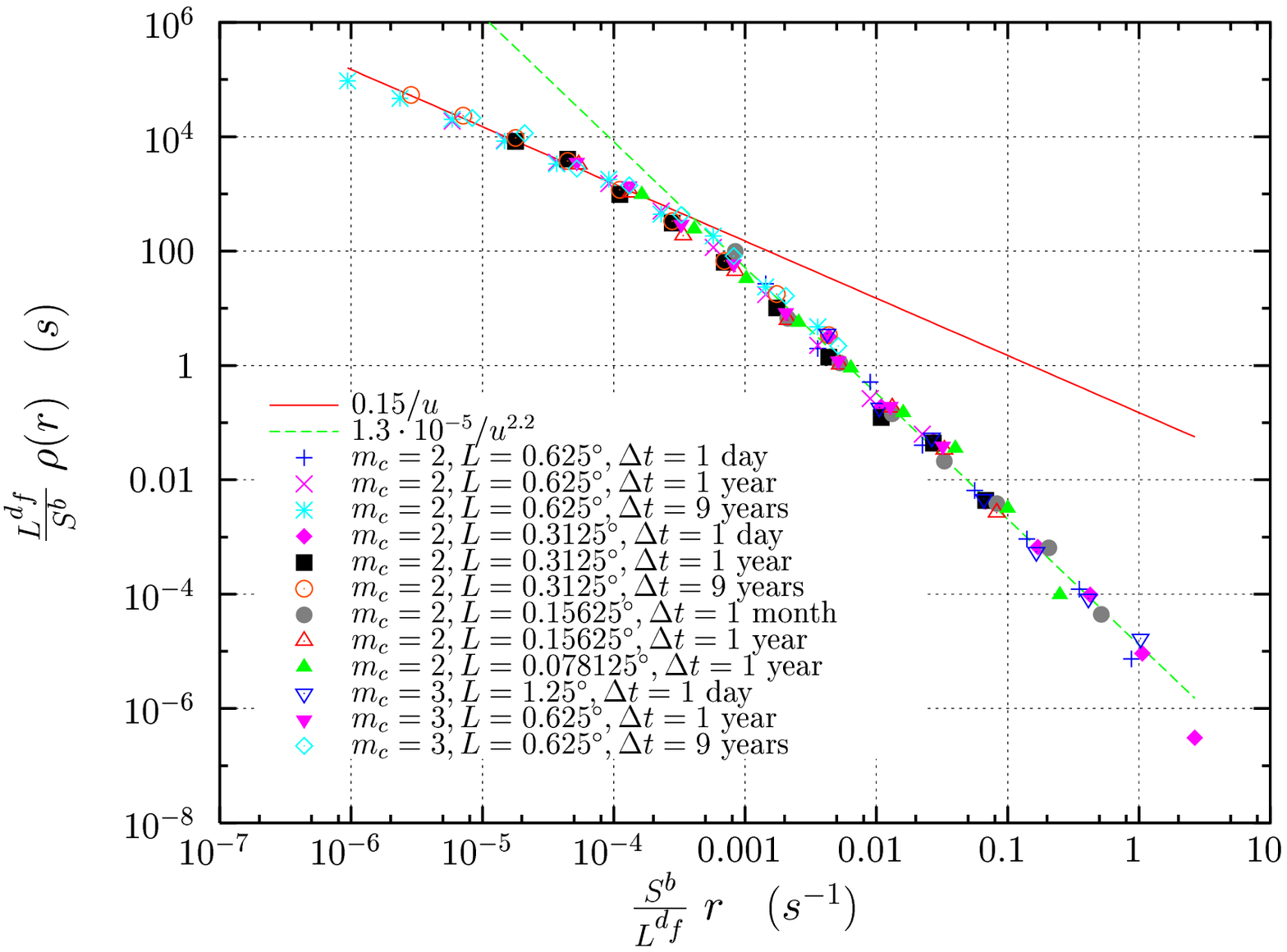} 
\caption{ 
Scaled distributions of rates, for several $\Delta t$, $L$, and $m_c$,
using $d_f=1.6$ and $b=0.95$.
Two power laws with exponents 1 and 2.2 fit the data.
\label{Drate}
}
\end{figure}

Next step is to look at the distribution of rates $\rho(r)$.
To be precise, $r(x,y,t)$ is defined by counting the number of events
above the threshold $m_c$ 
into the $L \times L -$region of coordinates $x,y$
during a time interval $(t,t+\Delta t)$ and
dividing the result by the duration of the interval, $\Delta t$.
The corresponding probability density, 
which is calculated from 1984 to 2001 
only for the regions in which there is earthquake activity,
depends on $L$ and $\Delta t$
(as well as on $m_c$) and is shown in Fig. \ref{Drate}.
In fact, the value of $\Delta t $ should be small enough to ensure
$r \simeq$constant, 
(but large enough for statistical significance),
but there is no characteristic scale for constant $r$
and therefore no typical value for $\Delta t$.
Also, it is noteworthy in the figure 
the two-power-law behavior, one power law for low rates 
and another one for high rates,
which can be modeled as
\begin{equation}
    \rho(r) =C' \theta \frac{(\theta r)^{\alpha-1}}
             {\left[1+(\theta r)^c\right]^\frac{\alpha+\beta}{c}},
\label{rho}
\end{equation}
which gives $\rho \propto 1/r^{1-\alpha}$ for $r \ll \theta^{-1}$
and $\rho \propto 1/r^{1+\beta}$ for $r \gg \theta^{-1}$.
We obtain exponents for $r$ about $1$ and $2.2$, so $\alpha \simeq 0$ and
$\beta \simeq 1.2$.
Parameter $c$ just controls the sharpness of the transition
from one regime to the other
and $\theta^{-1}$ is a scaling parameter.
The fact of having two power laws means that there is no characteristic
occurrence rate up to the value $\theta^{-1}$, but for values in the
tail of the distribution there is also scale invariance.
This could be understood as criticality, not only in the time domain
as we knew, but also in the rate domain.

Additionally, it is easy to obtain the form of the scaling
factor $\theta^{-1}$. 
The mean rate $\mu=\langle r \rangle$ is given
by the total number of events divided 
by the total time 
and by the number of regions with activity; 
the former, because of the Gutenberg-Richter law scales
as $1/S^b$ and the latter as $1/L^{d_f}$, which gives $\langle r \rangle
\propto L^{d_f}/S^b $.
Since the distribution turns out to scale in the same 
way, $ \theta^{-1} \propto L^{d_f}/S^b $.
For the scaling plot in Fig. \ref{Drate} 
we have used 
$b=0.95$ \cite{BCDS} and
$d_f= 1.6$,
which was obtained from a box-counting method for 
spatial distributions of epicenters with $m \ge 2$.


Now that we know the form of the functions $D(\tau|r)$ and $\rho(r)$
we can answer the question about how the large variations of the rate
influence the distribution $D(\tau)$, just by integrating Eq. (\ref{int})
with the use of (\ref{conditional})-(\ref{rho}).
The limit $\tau \rightarrow \infty$ is obtained directly with the use
of Laplace's method to evaluate asymptotic integrals \cite{Bender}.
We get
\begin{equation}
   D(\tau) \propto \frac{\theta^\alpha}{\mu} \tau^{\gamma-1}
        \int_0^{\infty} r^{\alpha+\gamma} e^{-(\tau r/u_0)^\delta} dr \\
        \propto 
        \frac{\theta^\alpha}{\delta \mu \tau^{2+\alpha}};
\end{equation}
this is in fact independent on the tail of $\rho(r)$,
it does not matter if it is a power law or not.
But there is another limit to be studied.
Indeed, the behavior of the integral depends on the relation 
between $\tau$ and $\theta$.
We have just calculated what happens for $\tau \gg \theta$;
the opposite case, $\tau \ll \theta$, can be obtained for long times
if we first perform the limit $\theta \rightarrow \infty$ 
and then apply Laplace's method for $\tau$.
So, $\rho(r) \sim C' \theta^{-\beta} / r^{1+\beta}$, and the integral gives
\begin{equation}
   D(\tau) \propto \frac{\tau^{\gamma-1}}{\mu \theta^\beta} 
        \int_0^{\infty} r^{\gamma-\beta} e^{-(\tau r/u_0)^\delta} dr \\
        \propto
        \frac{1}{\delta \mu \theta^\beta \tau^{2-\beta}}.
\end{equation}

\begin{figure}
\epsfxsize=3.5truein \hskip 1truein\epsffile{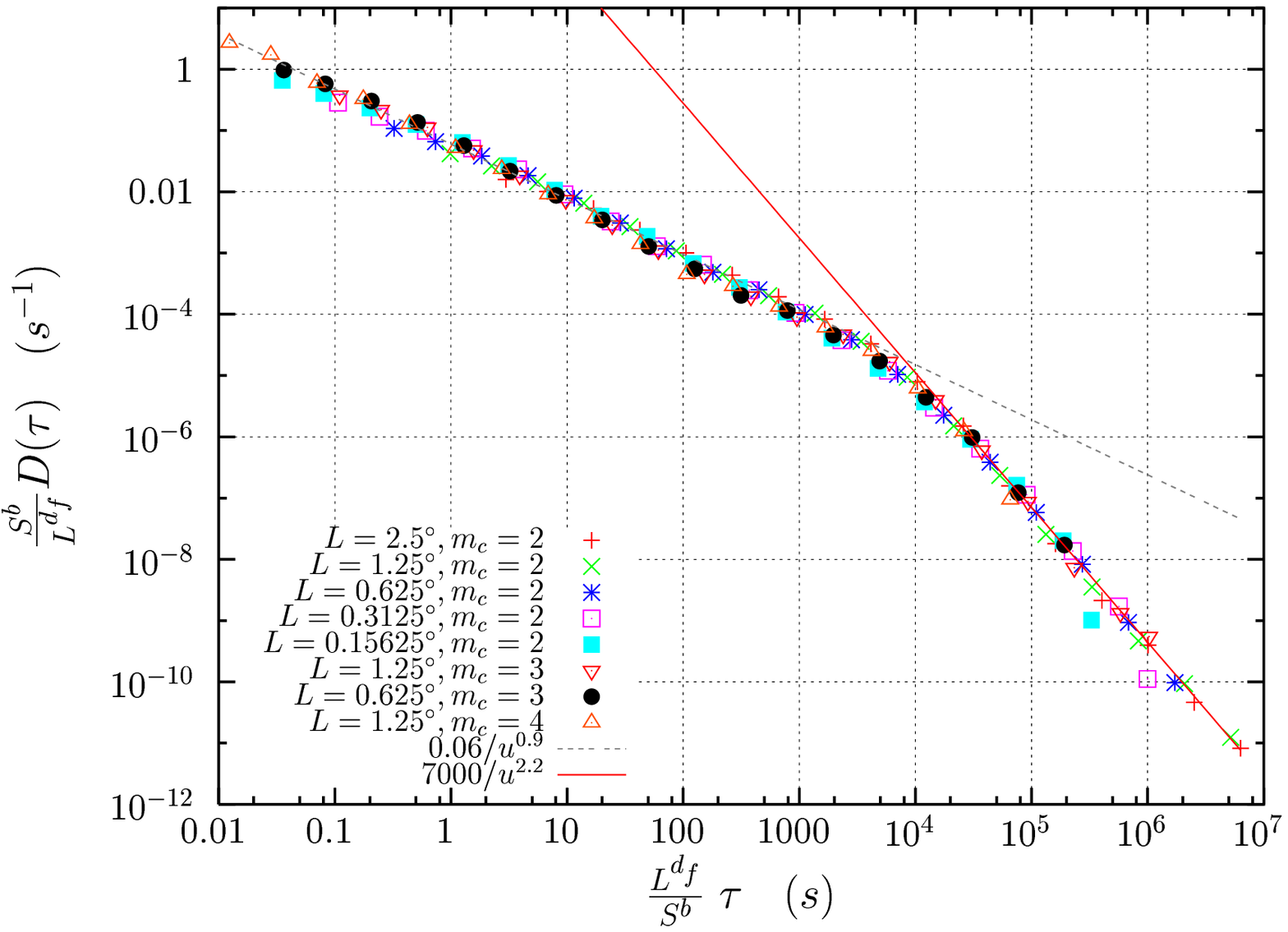} 
\caption{ 
Scaled distributions of interoccurrence times, $D(\tau)$,
for different $L$ and $m_c$ with $d_f=1.6$ and $b=0.95$;
$\tau \ge 38 s$ again.
The straight lines illustrate the double power-law behavior,
with exponents 0.9 and 2.2.
\label{Dt}
}
\end{figure}

Since $\mu$ is proportional to $\theta^{-1}$
the last results can be summarized as follows,
\begin{equation}
    D(\tau) \propto \frac{\theta^{1-\beta}} {\tau^{2-\beta}}
    \mbox{ for } \tau \ll \theta,
    \ \mbox{ and } \
    D(\tau) \propto \frac{\theta^{1+\alpha}} {\tau^{2+\alpha}}
    \mbox{ for } \tau \gg \theta,
\end{equation}
and are displayed in Fig. \ref{Dt}, 
where the two exponents for $\tau$ turn out to be
about 0.9 and 2.2, giving $\beta \simeq 1.1$ and $\alpha \simeq 0.2$ 
in good agreement with our calculations.
Both results for $D(\tau)$ do not depend 
on the form given by Eq. (\ref{rho}) for $\rho(r)$
as long as it exhibits the two-power-law behavior.
In addition, the exponent $\gamma$ of $D(\tau|r)$ does not affect
the value of the two exponents of $D(\tau)$.
Notice that 
the scaling factor for $\tau$ in $D(\tau)$ is $\theta$,
that is, the inverse of the scaling factor for $r$,
so, $\theta \propto S^b/L^{d_f}$.

%

Finally, 
we would like to point out
that $D(\tau)$ is described by the same function 
(including the same values of the exponents) 
than the one that characterizes the trapping time distribution
in a rice-pile model, see Fig. 1 of Ref. \cite{Boguna}.
Also the coincidence between our Fig. \ref{Drate}
and Fig. 3 in Ref.  \cite{Corral99} is notable,
although no probability density is measured there.
The exponents about 0.9 and 2.2 should be quite universal.

In conclusion, we have performed a "microscopic" analysis of the BCDS
law for earthquakes,
which provides a way to deal with the heterogeneity and
non-stationarity of seismic occurrence.

The author is profoundly indebted to Per Bak,
who opened so many paths in Science,
not only for his scientific guide,
but for his personal warmth as well.
Regarding this paper, he also thanks M. Bogu\~n\'a,
K. Christensen, and Ram\'on y Cajal program.
A fruitful part of this work was accomplished at l'Abadia de Burch 
(Pallars Subir\`a, Lleida).

%
%

%
%

%

%
%



\begin{references}
\bibitem{L}
Note that, due to the use of the longitude, 
the sides of a region do not have the same length 
and the regions are not equally sized .
Nevertheless this does not seem to be a problem at all.
\bibitem{variability}
In fact, it is an infinite variability,
due to its power-law distribution, as we will see.
\end{references}
\end{document}